\documentclass[briefpaper,9pt]{IEEEtran}
\usepackage{cite}
\usepackage{amsmath,amssymb,amsfonts}
\usepackage{algorithmic}
\usepackage{graphicx}
\usepackage{textcomp}
\usepackage{xcolor}
\usepackage{mathrsfs}
\usepackage{subfigure}
\usepackage{amsfonts}
\usepackage{cite}

\usepackage{indentfirst}

\usepackage{flushend}

\usepackage{multirow}
\usepackage{float}
\usepackage{algorithm}
\usepackage{algorithmic}
\usepackage{nomencl}

\usepackage{etoolbox}

\allowdisplaybreaks

\begin{document}

\begin{@twocolumnfalse}

    \title{
    \vspace{0.5cm}
        \large \textbf{Unified Predefined-time Stability Conditions of Nonlinear Systems with Lyapunov Analysis }}
    \author{
    	Bing Xiao, Haichao Zhang, Shijie Zhao, and Lu Cao
\thanks{Bing Xiao, Haichao Zhang, and Shijie Zhao are with the School of Automation, Northwestern Polytechnical University, Xi'an 710072, China. Lu Cao is with the National Innovation Institute of Defense Technology, Chinese Academy of Military Science, Beijing 100071, China. (Email: xiaobing@nwpu.edu.cn; zhanghaichao@mail.nwpu.edu.cn; zhaoshijie@mail.nwpu.edu.cn; caolu\_space2015@163.com). }
\thanks{    {Corresponding author: Bing Xiao.}}
}
    \maketitle
\end{@twocolumnfalse}


{\bf Abstract:} This brief gives a set of unified Lyapunov stability conditions to guarantee the predefined-time/finite-time stability of a  dynamical systems. The derived Lyapunov theorem for autonomous systems establishes equivalence with existing theorems on predefined-time/finite-time stability. The findings proposed herein develop a  nonsingular sliding mode control framework for an Euler-Lagrange system to analyze its stability, and its upper bound for the settling time can be arbitrarily determined a priori through predefined time constant. 

{\bf Introduction:} As we known, the convergence rate constitutes a key metric for evaluating system control performance. The finite-time control method was introduced to attain rapid stability \cite{bhat_geometric_2005}, thereby addressing limitations inherent in asymptotically stable control systems characterized by infinite convergence time. Its convergence time is contingent upon the system's initial states, resulting in a variable settling time corresponding to different initial states, and becomes indeterminate. The fixed-time control method came about to make the upper bound on the convergence time less conservative \cite{andrieu_homogeneous_2008,polyakov_nonlinear_2012}. It severs the dependence of settling time on the system's initial states. However, it involves a complex function with multiple control gains to characterize the convergent time upper bound, making it challenging to arbitrarily select a convergence time through the tuning of control gains. To this end, the predefined-time control technique with the settling time determined by only a parameter was discussed in \cite{Sanchez-Torres_2018_ClassPredefinedtimea,Aldana-Lopez_2019_EnhancingSettling,Jimenez_2020_Lyapunov,Kewen_2023_AdaptivePredefinedTime}. This approach has greater flexibility in determining the settling time than finite-time and fixed-time control methodologies.

Lyapunov theory, an effective instrument for analyzing the stability of control systems, is often combined with sliding mode control, backstepping control, and adding power integrator technique to design controllers to ensure the closed-loop control system's finite-time \cite{feng_non-singular_2002}, fixed-time \cite{zuo_nonsingular_2015,Zou_2020_FixedtimeAttitude}, and predefined-time stability \cite{Jang_2023_Predefinedtime,Xiao_2023_Prescribed}. Particularly, the sliding mode controller is usually closed-loop in the control system to obtain predefined-time stability \cite{Vazquez_2019_PredefinedtimeRobust,Ye_2022_Predefinedtime,Xiao_2023_Prescribed}. As detailed in \cite{Wang_2020_AttitudeControla,Xu_2021_DistributedPrescribedtime,Wu_2021_PredefinedtimeAttitude}, some specific instantiations of the predefined-time controller were formulated to achieve the control requirements of second-order systems featuring uncertainties,  but challenges related to singularity were encountered. To overcome this mentioned drawback,  the alternative nonsingular predefined-time sliding mode  controllers were presented in \cite{Ye_2022_Predefinedtime,Xiao_2023_Prescribed,Liang_2021_NovelSlidingb}. The aforementioned controllers, designed using different Lyapunov-based theorems, are applied to many systems and exhibit predefined-time stability properties. In fact, by analyzing them, a unified Lyapunov theorem that covers  existing Lyapunov theorems can be derived and ensures autonomous systems's predefined-time stability equally.

This brief advances unified sufficient conditions for the predefined-time/finite-time stability of autonomous systems using Lyapunov theory. The main contributions are stated as follows: a unified Lyapunov theorem is proposed to guarantee a class of dynamic systems to exhibit predefined-time stability. It serves to consolidate and unify the previous contributions of Lyapunov-based predefined-time stability theorems for autonomous systems in the literature \cite{Aldana-Lopez_2019_EnhancingSettling,Jimenez_2020_Lyapunov,Liang_2021_NovelSlidingb,Ni_2021_PredefinedtimeConsensus,Vazquez_2019_PredefinedtimeRobust}. Particularly, not like the results in \cite{Jimenez_2020_Lyapunov}, which found that the predefined-time Lyapunov theorem could only be developed on the basis of a  $\mathcal{K}_1$ function. The results presented in this brief relax this constraint, and an arbitrary strictly monotonically bounded increasing or decreasing function can be allowed. Moreover, if the selected function is a strictly monotonically unbounded increasing or decreasing function, the predefined-time stability conclusion will degenerate into a finite-time stable solution. Hence, the finite-time stable corollary is derived, and it can cover the existing Lyapunov finite-time stability theorems in \cite{Bhat_2000_FinitetimeStability,Chen_2010_FinitetimeStability,Shen_2009_UniformlyObservable}. The proposed unified Lyapunov theorem is also employed in the design of a suite of nonsingular sliding mode controllers for Euler-Lagrange system, ensuring its predefined-time stability.  Subsequently, the efficacy of the aforementioned properties is substantiated by Monte Carlo simulation examples. It provides a comprehensive exposition of the proposed controller's behavior, specifically focusing on settling time illustration.

{\bf Notation:} Throughout the article, the following notation are used. $\mathfrak{R}$ is a set of real numbers. For $\boldsymbol{\chi} \in \mathfrak{R}^n$, $\boldsymbol{\chi}^{\top }$ denotes its transpose,  $\dot{\boldsymbol{\chi}}=\frac{d \boldsymbol{\chi}}{d t}$ denotes the time derivative  of $\boldsymbol{\chi}$. For a scalar function $h(\boldsymbol{\chi}): \mathfrak{R}^{n} \rightarrow \mathfrak{R}$, $e^{h(\boldsymbol{x})}$ represents a standard exponential function with a natural constant $e$ as the base; $|h(\boldsymbol{\chi})|$ denotes the absolute value $h(\boldsymbol{\chi})$. A function $P(\chi):  [0,\infty) \rightarrow[0, a)$ is said to be a class $\mathcal{K}_a$ function, if it is strictly increasing with $P(0)=0$.

 \textbf{Problem Statement:}
Consider the following autonomous system
 \begin{equation}\label{Eq1}
 	\dot{\boldsymbol{x}}=\boldsymbol{f}(\boldsymbol{x} ; \boldsymbol{u})
 \end{equation}
where $\boldsymbol{x}\in \mathfrak{R}^{n}$ is the system state and  $\boldsymbol{u} \in \mathfrak{R}^{m}$ stands for the control input. $\boldsymbol{x}_{0}$ is the system initial state. The origin $\boldsymbol{x}\equiv \mathbf{0}$ is the unique equilibrium of system \eqref{Eq1}. $\boldsymbol{f}: \mathfrak{R}^{n} \times \mathfrak{R}^{m} \rightarrow \mathfrak{R}^{n}$ denotes a smooth vector field function.
The objective is to give a unified Lyapunov stability theorem to guarantee the predefined-time stability of the system \eqref{Eq1}. Further, with the application of this theorem, design a nonsingular sliding mode control framework to guarantee an Euler-Lagrange system's predefined-time stability.

\textbf{Motivating example:} 
A  predefined-time stable system is designed as $\dot x=-\frac{1}{2 p T_c }e^{V^p} V^{-p} x$ with the Lyapunov function $V=\frac{1}{2} x^2$, $0<p<1$, and $T_c>0$ being the predefined time constant. Taking the time derivative of $V$ satisfies $\frac{d V}{dt} = -\frac{ 1}{ p T_c }e^{V^p} V^{1-p} <0, \; \forall {x}\neq  0$.
Intuitively, the system is asymptotically stable, due to the fact  $e^{V^p}>0$  always holds. Then, we  compute its convergence time. The time derivative of $V$ is rewritten as 	$\frac{d \psi(V)}{d t}  =-\frac{ 1}{T_c}$, where $\psi(V)=2-e^{-V^p}$ is a bounded increasing function with $\psi(V)\in [1,2)$.  Unlike literature \cite{Jimenez_2020_Lyapunov},  the strict constraint of the function $\psi(V)\in [0,\,1)$ being a $\mathcal{K}_1$ function is required. Therefore, $\psi(V)$ will decrease to the minimum $\psi_T=1$ from its arbitrary initial value $\psi(V_0)<2$. Then, integrating both sides $ {d \psi(V)}  =-\frac{ 1}{T_c} {d t}$ yields  $\int_{\psi(V_0)}^{\psi_T} {1}  d \psi= -\int_0^T \frac{1}{T_c} d t$, and simplifying it, one obtains $T = {\left(\psi(V_0)-\psi_T\right) T_c} < { T_c} $. It is proved  that $T\geq T_c $ is valid for any $\psi(V_0)$ decreases to the minimum of $\psi(V)$.  The Lyapunov candidate $V$  also converges to zero  simultaneously. The selected  $V$ is radially unbounded.  For any initial system state $x_0$, it can converge to the equilibrium  when $T\geq T_c$.

\textbf{Main Results: } Summarizing from the above motivating example, let the Lyapunov function $V$ be the independent variable of the strictly monotonically increasing bounded function $\psi(V)$, i.e., $\psi(V_0)<b$ with $b\in \mathfrak{R}$. Utilizing the characteristics of the monotonic function $\psi(V)$ forces $V$ to decay to zero along with the $\psi(V)$ function decreasing to its minimum. Therefore, we can establish an equivalent relationship between the system's stability and the decreasing  characteristic of function $\psi(V)$, and further reflect the system's settling time  through the decay time of function $\psi(V)$.  Inspired by this, we conclude unified sufficient Lyapunov conditions that aim to ensure a class of dynamic systems exhibits predefined-time stability.

{\it Theorem 1:}	For the system \eqref{Eq1}, there is  a  function $\psi(V)$ with $V$ being a positive, radically unbounded function,  and the following three sufficient conditions are satisfied:\\
(i) $\forall  { \boldsymbol x} \in \mathfrak{R}^n$, $ \psi(V)  \in[a, b)$, $\psi(0)=a$ with $a \in\mathfrak{R} $ and $b \in\mathfrak{R} $;\\
(ii) $\forall  { \boldsymbol x} \in \mathfrak{R} ^n$, $\frac{d \psi}{d\left(V^p\right)}>0$  with $0<p<1$;\\
(iii) $\forall  {\boldsymbol x}\neq  \boldsymbol 0$, $\frac{d V}{dt} \leq-\frac{b-a}{\frac{d \psi}{d\left(V^p \right)}} \frac{V^{1-p} }{p T_c}$ with $T_c>0$.\\
Then, the system \eqref{Eq1} is predefined-time stable, and its upper bound of the convergence time is $T_c$.

{\it Proof:}
Recalling the system $\eqref{Eq1}$ and the candidate Lyapunov selected as $V=\frac{1}{2} \boldsymbol x^\top \boldsymbol x$, if the designed control input can make the closed-loop system such that the sufficient condition (iii) in Theorem 1. The straightforward result that the system's origin is Lyapunov asymptotic stable can be obtained. 
Next, taking the time derivative of $\psi$ and invoking the sufficient condition (iii) yields
\begin{equation}\label{Eq6}
	\frac{d \psi}{d t}=\frac{d \psi}{d V} \frac{d V}{d t}=p V^{p-1} \frac{d \psi}{d\left(V^p\right)} \frac{d V}{d t} \leq-\frac{b-a}{T_c}
\end{equation}
From \eqref{Eq6}, it means the function $\psi(V)$ will stabilize at its minimum $\psi_T=\psi(0)=a$ from arbitrary initial value $\psi_0= \psi(V_0)< b$.  Therefore, the selected $V$ also decreases to zero simultaneously.  Then, calculate the convergence time of $\psi(V)$ from the initial value $\psi_0$ to its minimum $\psi_T$. Integrating both sides  $ {d \psi} \leq-\frac{b-a}{T_c} {d t}$ with respect to time, one has  $	\int_{\psi_0}^{\psi_T} d \psi \leq-\int_0^T \frac{b-a}{T_c} d t$.
As a consequence,  the inequality $T \leq\frac{\left(\psi_0-\psi_T\right) T_c}{b-a}=\frac{\left(\psi_0-a\right) T_c}{b-a}<T_c$ holds. 
Hence, $\psi(V)$ decreases from an arbitrary initial value $\psi_0$ to its minimum $\psi_T$ when $T \geq T_c$. Through the introduction of the function $\psi(V)$, as it undergoes a decrease from its initial value to the minimum within the predefined time $T_c$, its independent variable $V$ synchronously converges to the origin. Consequently, the state of system \eqref{Eq1}, starting from any arbitrary initial state $\boldsymbol{x}_0$, exhibits convergence towards the equilibrium state $\boldsymbol{x}\equiv\boldsymbol{0}$ when $T\geq T_c$.

As in the above analysis, the selection of $\psi(V)$ is a key to obtaining Theorem 1 and achieving predefined-time stability. Moreover, the existing Lyapunov predefined-time stability theorems, as documented in the literature \cite{Vazquez_2019_PredefinedtimeRobust,Wang_2020_AttitudeControla,Ni_2021_PredefinedtimeConsensus}, can be regarded as a particular case of Theorem 1. Several examples are given in the following:


{\it{Example 1:}} A candidate regulator function $\psi(V)= b-e^{-\alpha V^p}$ is selected with $b\in \mathfrak{R}$, $\alpha>0$, and $0<p<1$. The fact $\psi(V)\in[b-1, b)$ always holds, it means $b-a=1$. Differentiating $\psi(V)$ with respect to $V^p$, one has $d \psi(V)/d V^p=\alpha e^{-\alpha V^p} $. Thus, the sufficient condition (iii) becomes
\begin{equation}\label{Eq8}
	\frac{d V}{dt}  \leq- \frac{b-a}{\alpha p T_c}e^{\alpha V^p} V^{1-p}, \; \forall  {\boldsymbol x}\neq  \boldsymbol 0
\end{equation}
using Theorem 1, the system is predefined-time stable with $T< T_c$. When select $\alpha=1$, formula \eqref{Eq8} reduces to $	\frac{d V}{dt}  \leq- \frac{1}{ p T_c}e^{ V^p} V^{1-p}$, $\forall  {\boldsymbol x}\neq   \boldsymbol 0$. Hence, this specific case   covers the result in \cite{Wang_2020_AttitudeControla}.

{\it{Example 2:}}  A particular selection of $\psi(V)=\arcsin(\tanh(V^p))$ is given with $0<p<1$. One has $\psi(V)\in[0, \frac{\pi}{2})$ and $b-a=\frac{\pi}{2}$. Differentiating $\psi(V)$ with respect to $V^p$ yields $d \psi(V)/d V^p=\sqrt{1-\tanh(V^p)}$. The sufficient condition (iii) is given as
\begin{equation}\label{Eq18}
	\frac{d V}{dt}  \leq-\frac{b-a}{p T_{c}} \cosh(V^p) V ^{1-p}, \; \forall  {\boldsymbol x}\neq \boldsymbol  0
\end{equation}
on the basis of Theorem 1, the system is stable within predefined time $T_c$. The inequality \eqref{Eq18} is rewritten as $\frac{d V}{dt}  \leq-\frac{\pi}{2 p T_{c}} \cosh(V^p) V^{1-p}$, $\forall  {\boldsymbol x}\neq  \boldsymbol 0$. It is identical to  the  results of \cite{Ni_2021_PredefinedtimeConsensus}.

{\it{Example 3:}}  Select a regulator function $\psi(V)=\arctan(\sqrt{\frac{\beta}{\alpha}}V^p)$  with $\alpha>0$, $\beta>0$, and $0<p<1$. One  concludes $\psi(V)\in[0, \frac{\pi}{2})$, i.e., $b-a=\frac{\pi}{2}$. Differentiating $\psi(V)$ with respect to $V^p$ yields $d \psi(V)/d V^p=\sqrt{\alpha \beta}/({\alpha +\beta V^{2p}})$. Thus, the sufficient condition (iii) then becomes
\begin{equation}\label{Eq9}
	\frac{d V}{dt}  \leq-\frac{b-a}{p\sqrt{\alpha \beta}T_{c}}\left(\alpha V^{1-p} +\beta V^{1+p}\right), \; \forall  {\boldsymbol x}\neq  \boldsymbol 0
\end{equation}
using Theorem 1, its system state converges to origin when  $T\geq T_c$. If $\alpha=1$ and $\beta=1$, the inequality \eqref{Eq9} reduces to $  \dot{V} \leq-\frac{\pi}{2p T_{c}}\left( V^{1-p}+ V^{1+p}\right)$, which is equivalent to Theorem 1 of \cite{Vazquez_2019_PredefinedtimeRobust}.  

{\it{Example 4:}} To achieve predefined-time stability of an autonomous system in \cite{Jimenez_2020_Lyapunov}, the inequality $\dot \psi(V) \leq - \frac{1}{(1-p)T_c}\psi^p(V) $ needs to be satisfied. It derives the settling time function $T=T_c (\psi(V_0))^{1-p}<T_c$. It requires a strict constraint that the function $\psi(V)\in [0,\,1)$ is a $\mathcal{K}_1$ function. This brief breaks this constraint; only an arbitrary increasing function $\psi(V)\in [a,\,b)$ is needed, increasing the selection flexibility of $\psi(V)$. As a result, the unified Lyapunov Theorem 1 is derived to ensure an autonomous system's predefined-time stability and covers the results in \cite{Jimenez_2020_Lyapunov}.

These predefined-time stable examples are covered by Theorem 1 with a specific bounded monotonically increasing function $\psi(V)$. 
In fact, if $\psi(V)$ is a bounded monotonically decreasing function, we can also derive a Lyapunov-based predefined-time corollary as follows:

{\it Corollary 1:} For the system \eqref{Eq1}, there is  a  function $\psi(V)$ with $V$ being a positive, radically unbounded function,  and the following three sufficient conditions are satisfied:\\
(i) $\forall  { \boldsymbol x} \in \mathfrak{R}^n$, $ \psi(V)  \in (a, b]$, $\psi(0)=b$  with $a \in\mathfrak{R} $ and $b \in\mathfrak{R} $;\\
(ii) $\forall  { \boldsymbol x} \in \mathfrak{R}^n $, $\frac{d \psi}{d\left(V^p\right)}<0$  with $0<p<1$;\\
(iii) $\forall  {\boldsymbol x}\neq  \boldsymbol 0$, $\frac{d V}{dt} \leq\frac{b-a}{\frac{d \psi}{d\left(V^p \right)}} \frac{V^{1-p} }{p T_c}$ with $T_c>0$.\\
Then,  the system \eqref{Eq1} is  predefined-time stable, and the upper bound of the convergence time is $T_c$.

{\it Proof:} Considering the system $\eqref{Eq1}$ and the function $\psi(V)$ with $V=\frac{1}{2} \boldsymbol x^\top \boldsymbol x$. If the designed control input makes the closed-loop system meet the sufficient condition (iii) in Corollary 1, one directly derives that the system's origin is asymptotically stable.
Then, taking the time derivative of $\psi(V)$ and invoking sufficient condition (iii) yields $\frac{d \psi}{d t} \geq\frac{b-a}{T_c}$, it means that $\psi(V)$ will stabilize at its maximum $\psi_T=\psi(0)=b$ from arbitrary initial value $\psi_0= \psi(V_0)> a$.  Therefore, the selected $V$ also decreases to zero simultaneously.  Then, calculate the convergence time of  $\psi(V)$ from its initial value $\psi_0$ to its maximum $\psi_T $. Integrating both sides  $ {d \psi} \geq\frac{b-a}{T_c} {d t}$, one has $\int_{\psi_0}^{\psi_T} 1 d \psi \geq\int_0^T \frac{b-a}{T_c} d t$. As a consequence it holds: $	T \leq-\frac{\left(\psi_0-\psi_T\right) T_c}{b-a}=\frac{\left(b-\psi_0\right) T_c}{b-a}< T_c$.
Hence, $\psi(V)$ from arbitrary initial value $\psi_0$ to the maximum $\psi_T$, whose convergent time such that  $T \leq T_c$.  When the function $\psi(V)$ increases from the initial value to the maximum value within predefined time $T_c$, its independent variable $V$ also converges synchronously to the origin. Therefore, the arbitrary initial state $\boldsymbol x_0$ of system  \eqref{Eq1} will converge to the equilibrium $\boldsymbol x\equiv\boldsymbol 0$ when $T > T_c$.


%
%
%
%

The selection of  $\psi$ in Theorem 1 and Corollary 1 is key to ensure predefined-time stability. Some examples are listed  in Table \ref{tab1}:
\begin{table}[!h] 
	\center
	\caption{Some examples of candidate function}
	\label{tab1}
	\renewcommand\arraystretch{1.6}
	{\begin{tabular}{ c l} 
			\hline\hline
			Number & Functions \\ \hline
			1	    	&    $\psi=\frac{qV^p}{ {V^p+k}}$, $0<p<1$, $q>0$,  $k>0$              \\ \hline
			2	    	&   $\psi= m\tanh\left( n V^p\right)$, $0<p<1$, $n>0$, $m>0$                \\ \hline
			3	    	&   $\psi= q+e^{-\alpha V^p} $, $0<p<1$, $\alpha>0$, $q>0$             \\ \hline
			4	    	&  $\psi=  \frac{n}{ {m+e^{\alpha V^p}}} $, $0<p<1$, $\alpha>0$, $m>0$, $n>0$                \\ \hline
			5	    	&    $\psi=\frac{1}{ {V^p+k}}$, $0<p<1$, $k>0$               \\ \hline
			$\cdots$	    	&   $\cdots$   \\\hline  \hline            
	\end{tabular}}
\end{table}

%
%
%
%
%
%
%

	
Moreover, if the function $\psi(V)$ does not meet the condition (i) of Theorem 1, i.e., $\psi(V)\in [a, +\infty)$ is a strictly monotonically increasing unbounded function, the conclusion of Theorem 1 degenerates into finite-time stability. Its settling time can also be computed. Now,
differentiating $\psi(V)$ and invoking the condition  $\frac{d V}{dt} \leq-\frac{1}{\frac{d \psi}{d\left(V^p \right)}} \frac{V^{1-p} }{p T_c}$ with $T_c>0$ and $0<p<1$, one has 
$\frac{d \psi(V)}{d t} \leq-\frac{ 1}{T_c}$.
Therefore, $\psi(V)$ will decrease to the minimum $\psi_T=a$ from the initial value $\psi(V_0)$. Integrating it on both sides yields $\int_{\psi(V_0)}^{\psi_T} {1}    d \psi\leq  -\int_0^T \frac{1}{T_c} d t$. Then, one has $T \leq {\left(\psi(V_0)-a\right) T_c}  $. It is proved that for any $\psi(V_0)$, it decreases to the minimum of $\psi(V)$ when $T>{\left(\psi(V_0)-a\right) T_c}$.  The Lyapunov candidate $V$  also converges to zero  simultaneously. The selected  $V$ is radially unbounded.  For any initial system state $x_0$, it can converge to the equilibrium $ x\equiv0$ when $T> {\left(\psi(V_0)-a\right) T_c}$.  Therefore, we can summarize the following finite-time stability corollaries on the basis of Theorem 1 and Corollary 1.
	
{\it Corollary 2:}	For the system \eqref{Eq1}, there is  a    function $\psi(V)$ with $V$ being a positive, radically unbounded function,  and the following three sufficient conditions are satisfied:\\
	(i) $\forall  { \boldsymbol x} \in \mathfrak{R}^n$, $ \psi(V)  \in[a,+ \infty)$, $\psi(0)=a$  with $a \in\mathfrak{R} $;\\
	(ii) $\forall  { \boldsymbol x} \in \mathfrak{R} ^n$, $\frac{d \psi}{d\left(V^p\right)}>0$  with $0<p<1$;\\
	(iii) $\forall  {\boldsymbol x}\neq  \boldsymbol 0$, $\frac{d V}{dt} \leq-\frac{1}{\frac{d \psi}{d\left(V^p \right)}} \frac{V^{1-p} }{p T_c}$ with $T_c>0$.\\
	Then,  the system \eqref{Eq1} is  finite-time stable, and the upper bound of the convergence time is ${\left(\psi(V_0)-a\right) T_c}$.
	
{\it Corollary 3:} For the system \eqref{Eq1}, there is  a    function $\psi(V)$ with $V$ being a positive, radically unbounded function,  and the following three sufficient conditions are satisfied:\\
	(i) $\forall  { \boldsymbol x} \in \mathfrak{R}^n$, $ \psi(V)  \in (- \infty, a]$, $\psi(0)=a$  with $a \in\mathfrak{R} $;\\
	(ii) $\forall  { \boldsymbol x} \in \mathfrak{R}^n $, $\frac{d \psi}{d\left(V^p\right)}<0$  with $0<p<1$;\\
	(iii) $\forall  {\boldsymbol x}\neq  \boldsymbol 0$, $\frac{d V}{dt} \leq\frac{1}{\frac{d \psi}{d\left(V^p \right)}} \frac{V^{1-p} }{p T_c}$ with $T_c>0$.\\
	Then,  the system \eqref{Eq1} is  finite-time stable, and the upper bound of the convergence time is ${\left(a-\psi(V_0)\right) T_c}$.

\textbf{Predefined-time controller design: }Consider the following Euler-Lagrange system 
\begin{equation}\label{Eq11}
	\left\{	\begin{aligned}
		& \dot{x}_1=x_2\\
		& \dot{x}_2= f(x_1,x_2) +g(x_1,x_2)u  +w
	\end{aligned}\right.
\end{equation}
where $\boldsymbol{x}=\left[x_1\; x_2\right]^\top \in \mathfrak{R}^2$ is the available state vector; $g(x_1,x_2)\neq0$ and  $f(x_1,x_2)$  are known nonlinear functions; $w$ is the disturbances and system uncertainties. It satisfies $|w|\leq \kappa$, with $\kappa$ being a known positive scalar. $u\in \mathfrak{R}$ is the control input, which will be designed to guarantee the state trajectory of system \eqref{Eq11} is predefined-time stable.

Choose a function $\psi_1\in [a,b)$  satisfies the sufficient conditions (i) and (ii) of Theorem 1, and define $V_1=\frac{1}{2}  x_1^2 $. 
For the system \eqref{Eq11},  a class of nonsingular sliding mode surfaces is designed as
\begin{equation}\label{Eq13}
	\left\{\begin{aligned}
		 	s=&{  x}_2+ \frac{(b-a) V_1^q \Phi}{2p_1T_1}\\
		\Phi=&	\left\{ \begin{array}{l}
			H_1 x_1  V_1^{-p_1-q}\mbox{, } \qquad \qquad	V_1\geq \eta_0   \\
			H_1 x_1  (k_1 V_1+k_2 V_1^2)\mbox{,}\, \;\;\; \;V_1<\eta_0 
		\end{array}\right.  
	\end{aligned}\right.
\end{equation}
with  $0<p_1<\frac{1}{2}$, $q>1$, $\eta_0>0$, and $H_1 =1/\frac{d\psi_1(V_1)}{dV_1^{p_1}}$. $T_1$ is the predefined time constant.   $k_1=2\eta_0^{-1-p_1-q}$ and  $k_2=-\eta_0^{-2-p_1-q}$ are selected to meet the continuity of $	\boldsymbol	s$.  Differentiating $s$ yields 
\begin{equation}\label{Eq14}
	\dot	s=\dot { x}_2+ \frac{ (b-a) (\dot \Phi V_1^q + q  \Phi V_1^{q-1})}{2p_1T_1} \\
\end{equation}
where $\dot \Phi$ can be computed by mathematical operations form  \eqref{Eq13}.  Specifically, the fractional order term that is prone to generating singular values appears in the term $H_1 x_1  V_1^{-p_1-q}$. It can be observed that when $V_1\geq \eta_0$, even if it appears, $\dot \Phi$ will not exhibit singularity. When $V_1<\eta_0$, there is no singularity, and the system's state will asymptotically converge to the origin.

For system \eqref{Eq11}, the control variable $u$ is designed as
\begin{equation}\label{Eq12}
	\begin{aligned}
		u= &g^{-1}(x_1,x_2) \Big(-  \frac{ (b\!-\!a) (\dot \Phi V_1^q +\! q  \Phi V_1^{q-1})}{2p_1T_1} \\
		& -\frac{(b\!-\!a)H_s s}{2 T_2 p_1}V_2^{-p_1}-\kappa {\rm sign}(s) -f(x_1,x_2)\Big  )
	\end{aligned}
\end{equation}
with $H_s =1/\frac{d\psi_1(V_2)}{dV_2^{p_1}}$ and $V_2=\frac{1}{2}s^2$ being  a Lyapunov function. $T_2$ is the convergence time of reaching phase predefined by user.

{\it Theorem 2:}  The system \eqref{Eq11} closed-loop the controller \eqref{Eq12} will reach the sliding mode surface \eqref{Eq13} within predefined time $T_2$, and the states $x_1$ and $x_2$ converge to a small region around zero along the surface within a predefined-time stable $T=T_1+T_2$.

 {\it Proof}:
Differentiating $V_2$ and inserting \eqref{Eq14} and \eqref{Eq12} yields
\begin{equation}\label{Eq15}
	\begin{aligned}
		\dot V_2=&s\Big( f(x_1,x_2) +g(x_1,x_2)u  +w -\frac{(b-a)H_s s}{2 T_2 p_1}V_2^{-p_1}\Big) \\
		\leq&-\frac{b-a}{2H_s T_2 p_1}V_2^{1-p_1} 
	\end{aligned}
\end{equation}
from \eqref{Eq15}, $V_2$ will converge to zero within predefined time $T_2$ on the basis of Theorem 1. The ideal sliding mode motion is established simultaneously. Once $s$ reaches zero,  the designed sliding surface \eqref{Eq13} such that $s =0$. Then, one has $ { x}_2=- \frac{ H_1  x_1}{2  p_1T_1}V_1^{-p_1}$, when $|x_1|\geq\sqrt{2\eta_0}$. Taking the time derivative of $V_1$, one can obtain $\dot V_1=- \frac{ H_1  }{  p_1T_c}V_1^{1-p_1}$.  Using Theorem 1,  $ x_{1}$ and  $ x_{2}$ will converge to origin along the  sliding manifold within predefined time $T_1$. When $|x_1|<\sqrt{2\eta_0}$ will approach zero along the general sliding manifold. One has $ { x}_2=- \frac{ H_1  x_1}{2  p_1T_1}(k_1 V_1+k_2 V_1^2)$. The second phase of $s$ in \eqref{Eq13} is asymptotically stable.  The predefined-time convergence of state $x_1$, i.e.,   $\lim_{t\to( T_1+T_2)}|x_1|<\sqrt{2\eta_0}$ is achieved, and the singularity problem of predefined-time sliding control	can be circumvented.

\textbf{Simulation examples:}
To validate the predefined-time stability of the system  \eqref{Eq11} closed-loop with the  controller \eqref{Eq12}, the Monte Carlo simulations with 500 dispersed scenarios are conducted. For simulation purposes, let $f(x_1,x_2) =x_1^2+x_1\sin(x_2)$ and $g(x_1,x_2)=(x_1+x_2)^2+1$; the bounded external disturbance is supposed to be $w= 0.1\sin(t)$. The initial conditions of $x_1$ and $x_2$ satisfy $x_1(0)\in[-1200, 1200]$ and $x_2(0)\in[-100, 100]$.
\begin{figure}[!h] 
	\centering
	\subfigure[Behavior of $x_1$]{\includegraphics[width=0.2\textwidth]{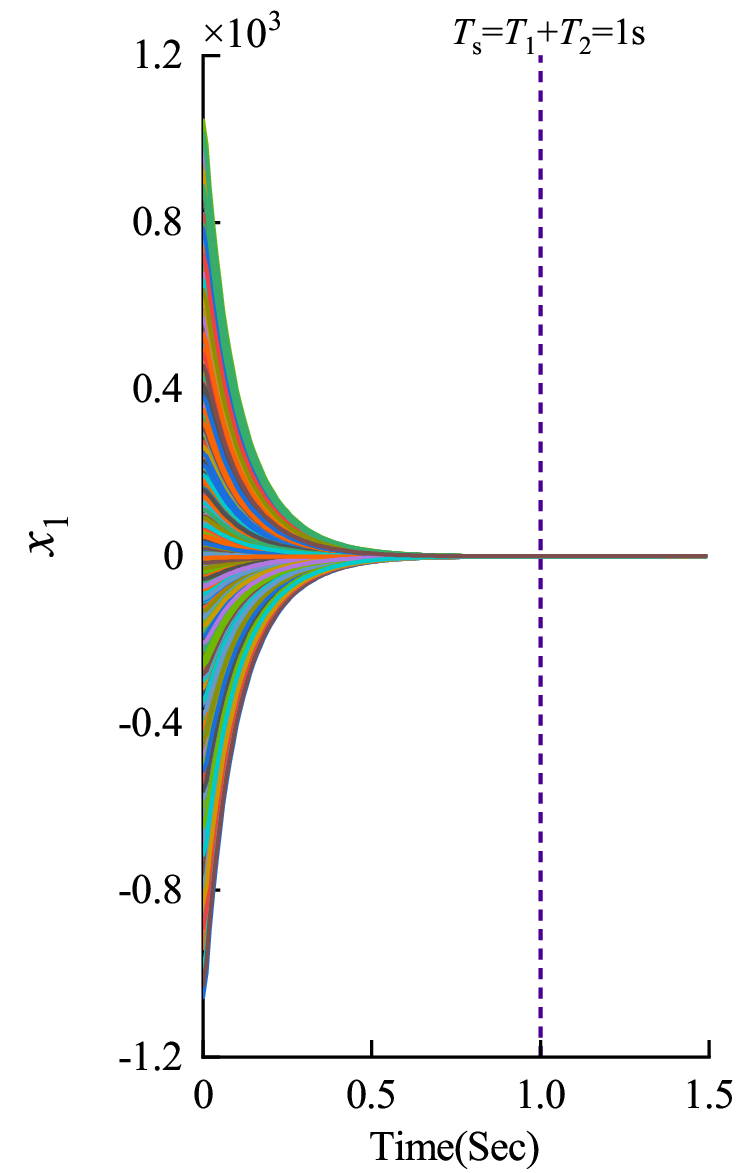}}
	\subfigure[Behavior of $s$ ]{\includegraphics[width=0.2\textwidth]{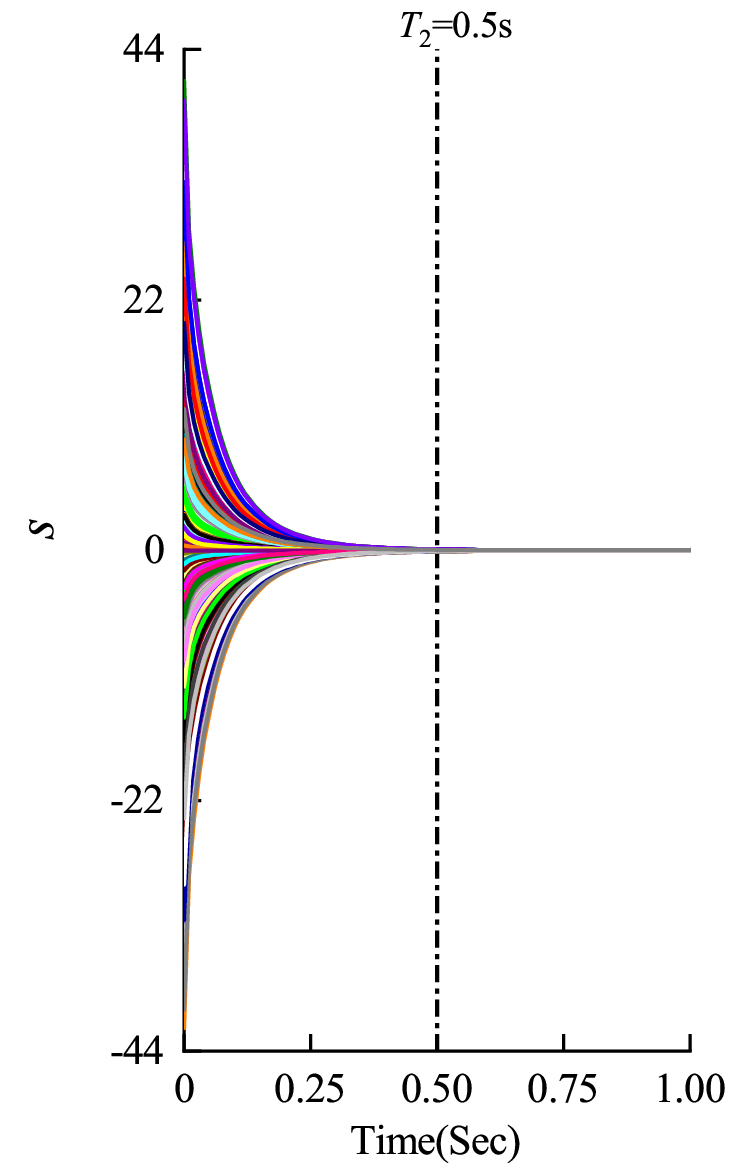}}
	\caption{Convergence behavior of sliding surface $s$ and state $x_1$  in the  Monte Carlo simulations with an increasing function $\psi_1(\nu)= {b}/({ {a+e^{-\alpha \nu^{p_1}}}}) $.}
	\label{fig1}
\end{figure}  

\begin{figure}[!h] 
	\centering
	\subfigure[Behavior of $x_1$]{\includegraphics[width=0.2\textwidth]{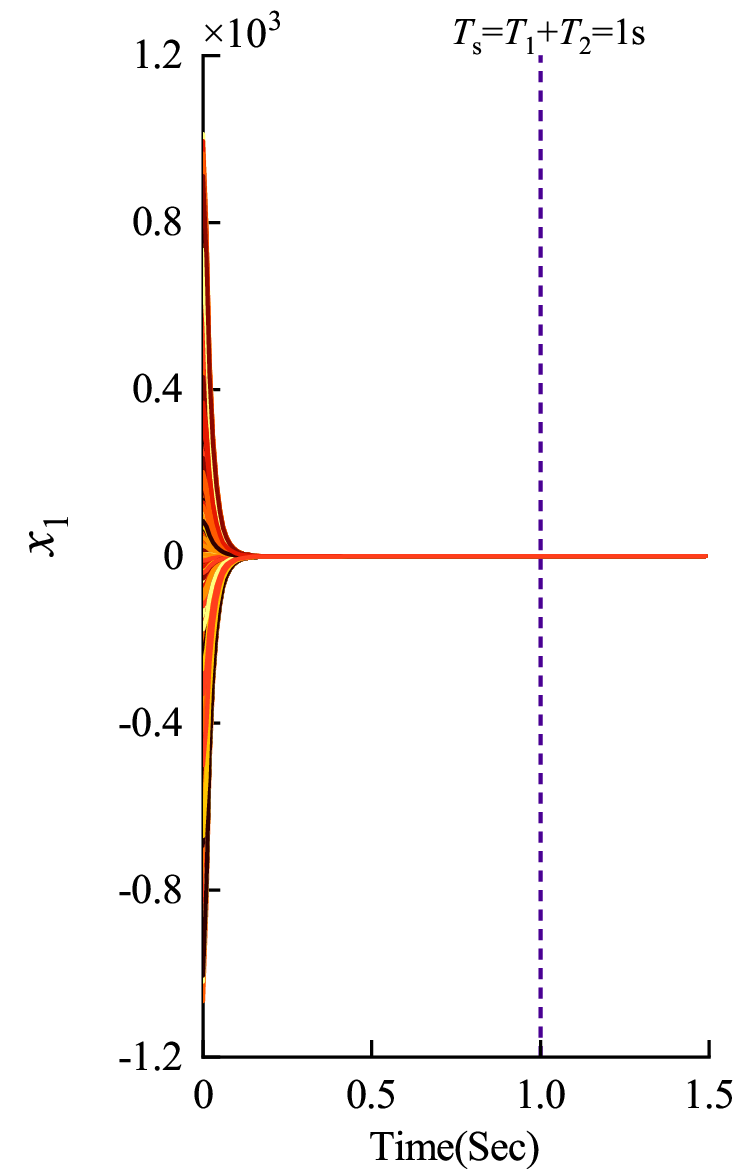}}
	\subfigure[Behavior of $s$ ]{\includegraphics[width=0.2\textwidth]{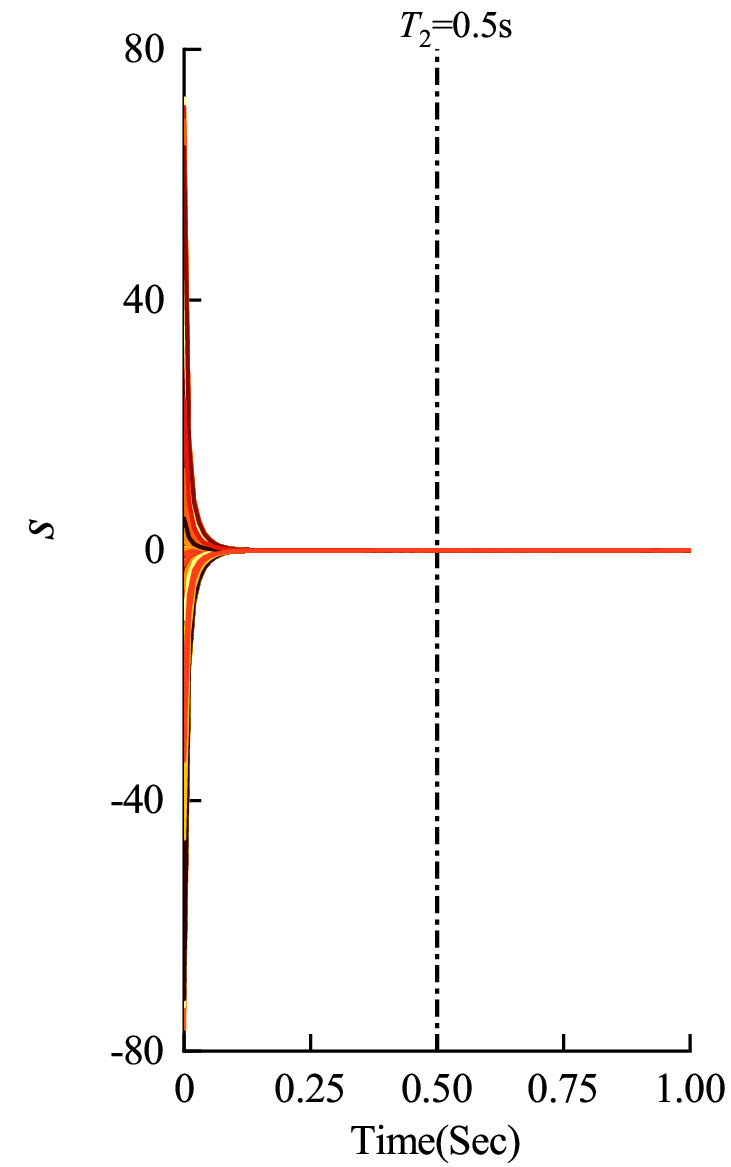}}
	\caption{Convergence behavior of sliding surface $s$ and state $x_1$ in the  Monte Carlo  simulations with a decreasing function $\psi_1(\nu)={\bar a+e^{-\bar \alpha \nu^{  p_1}}} $.}
	\label{fig2}
\end{figure}  
On the one hand, in sliding surface \eqref{Eq13}, we select a monotonically increasing  function   $\psi_1(\nu)= {b}/({ {a+e^{-\alpha \nu^{p_1}}}}) $ with $0<p_1<1$, $\alpha>0$, $a>0$, and $b>0$. One can obtain that $\psi_1(\nu)\in[\frac{b}{a+1}, \frac{b}{a})$. Therefore, one has $H_1=\frac{1}{b\alpha}(a+e^{-\alpha V_1^{p_1}})^2e^{\alpha V_1^{p_1}}$ and $H_s=\frac{1}{b\alpha}(a+e^{-\alpha V_2^{p_1}})^2e^{\alpha V_2^{p_1}}$.
The control input \eqref{Eq12}, guaranteeing predefined-time stability, has been tested by Monte Carlo simulations with the design choices $a=1$, $b=3$, $\alpha=1$, and  $p_1=0.051$.  The predefined time constants are  $T_1=0.5$ and $T_2=0.5$. Fig. \ref{fig1} displays the convergence behavior of the sliding surface  and the system state. It illustrates that the system state converges to the origin within the predefined time $T_s=T_1+T_2=1$s.

On the other hand, in sliding surface \eqref{Eq13}, we select a monotonically decreasing  function   $\psi_1(\nu)={\bar a+e^{-\bar \alpha \nu^{  p_1}}} $ with $0<  p_1<1$, $\bar \alpha>0$, and $\bar a>0$. One can obtain that $\psi_1(\nu)\in(\bar a, \bar a+1]$. Therefore, one has $H_1=-\frac{1}{\bar\alpha}e^{\bar \alpha V_1^{\bar p_1}}$ and $H_s=-\frac{1}{\bar\alpha}e^{\bar \alpha V_2^{\bar p_1}}$. The control input \eqref{Eq12}, ensuring predefined-time stability, has been checked by Monte Carlo simulations with the control gains  $\bar \alpha=1$ and  $p_1=0.05$.  The given time constants are  $T_1=0.5$ and $T_2=0.5$.  Fig. \ref{fig2} plots the convergence performance of sliding surface and system state. One can see the system state converges to the origin within the predefined time $T_s=T_1+T_2=1$s.

\textbf{Conclusion:} This brief has presented the sufficient conditions for predefined-time/finite-time stability of autonomous systems within the framework of Lyapunov theory. The developed Lyapunov-based results allow us to establish equivalence with existing Lyapunov-based theorems in predefined-time/finite-time stability for autonomous systems. On this basis, a nonsingular sliding mode control method is provided to analyze the predefined-time stability of an Euler-Lagrange system. The effectiveness and superiority of the proposed control approaches have been verified through the numerical simulation results.

\textbf{Acknowledgments:} This work was supported by the National Natural Science Foundation of China.

\bibliographystyle{IEEEtran}
\bibliography{reference.bib}

\end{document}